\documentclass[twocolumn,amssymb,amsfonts,amsmath,nofootinbib,aps,prb,%
reprint,superscriptaddress,letterpaper,showpacs,floatfix]{revtex4-1}
\usepackage{graphicx}
\usepackage{dcolumn}
\usepackage{bm}
\usepackage{mathtools}
\usepackage{mathrsfs}
\usepackage{color}
\newcommand{\imsize}{\columnwidth}
\begin{document}
\title[]{Density Matrix Embedding from Broken Symmetry
Lattice Mean-Fields}
\author{Ireneusz W. Bulik}
 \affiliation{Department of Chemistry, Rice University, Houston, Texas 77005, USA}
\author{Gustavo E. Scuseria}
 \affiliation{Department of Chemistry, Rice University, Houston, Texas 77005, USA}
 \affiliation{Department of Physics and Astronomy, Rice University, Houston, Texas 77005, USA}
\author{Jorge Dukelsky}
 \affiliation{Instituto de Estructura de la Materia-CSIC, Serrano 123, 28006 Madrid, Spain}
\date{\today}
\begin{abstract}
Several variants of the recently proposed Density Matrix Embedding Theory
(DMET) [G. Knizia and G. K-L. Chan, {\it Phys. Rev. Lett.} {\bf 109}, 186404
(2012)] are formulated and tested.  We show that spin symmetry breaking of the
lattice mean-field allows precise control of the lattice and fragment filling
while providing very good agreement between predicted properties and exact
results.  We present a rigorous proof that at convergence this method is
guaranteed to preserve lattice and fragment filling.  Differences arising from
fitting the fragment one-particle density matrix alone versus fitting
fragment plus bath are scrutinized.  We argue that it is important to restrict
the density matrix fitting to solely the fragment.  Furthermore, in the
proposed broken symmetry formalism, it is possible to substantially simplify
the embedding procedure without sacrificing its accuracy by resorting to
\emph{density} instead of \emph{density matrix} fitting.  This simplified
Density Embedding Theory (DET) greatly improves the convergence properties of
the algorithm.
\end{abstract}

\maketitle
\section{Introduction}

Strongly-correlated electron systems are currently of great interest in the
condensed matter physics and quantum chemistry communities.
\cite{annurev-physchem-032210-103338,
ADMA:ADMA201202018} Despite the differences, for
example, between frustrated lattices and chemical bond breaking, the
strong-correlation phenomenon in these systems has similar roots which are
manifested by the breakdown of the mean-field picture.  Regardless of the
details of the problem at hand, strong-correlation has significant impact on
many important aspects of the physics of various systems and cannot be
ignored. \cite{RevModPhys.66.763,ZPB.64.189,Science.309.257,Science.283.2034}
A low computational cost, qualitatively correct description of strongly-correlated
materials would have great impact on the quality of theoretical predictions.
Hence, the quest for developing novel, as well as improving existing
approaches for strong-correlation continues unabated.

One of the reasons that robust methods for the treatment of strong-correlation
are so elusive is related to the size of the systems of current interest.
With increasing number of electronic degrees of freedom,
the numerical complexity of exactly solving the problem quickly becomes
prohibitively large; approximate methods must be therefore employed.
An ideal approximation should provide a systematic and qualitatively
correct description that is computationally accessible and does not
deteriorate with system size. Such an ideal
tool would be therefore applicable to a broad range of problems.
From a practical point of view, approximate methods
should offer a good compromise between accuracy and computational
cost. Many methods are available to tackle the strong-correlation
problem; they include Quantum and Variational Monte Carlo,
\cite{PhysRevB.36.381,PhysRevB.71.241103,PhysRevB.85.045103,PhysRevLett.78.4486,
PhysRevLett.104.116402,PhysRevB.64.184509}
Dynamical Mean-Field Theory (DMFT), \cite{PhysRevLett.62.324,PhysRevLett.69.168,
PhysRevLett.69.1236,PhysRevLett.69.1240,PhysRevLett.72.1545,
PhysRevLett.104.226402,PhysRevB.78.115102,EPL.85.17002,PhysRevB.69.195105,
RevModPhys.68.13}
Density Matrix Renormalization Group (DMRG)
\cite{PhysRevLett.69.2863,RPP.67.513,RevModPhys.77.259, PhysRevB.53.R10445},
methods based on symmetry breaking and restoration,
\cite{PhysRevB.87.235129,JChemPhys.135.124108,JChemPhys.136.164109,
PhysRevB.85.245130,PhysRevB.87.115136,PhysRevB.69.045110}
and methods based on a Gutzwiller variational approach.
\cite{NewJPhys-11-075010-2009,PhysRevB.85.035133}
This list is not, by any means, exhaustive.
The effort to develop new approximations is continuously undertaken.

Recently, Knizia and Chan introduced Density Matrix Embedding
Theory (DMET), \cite{PhysRevLett.109.186404,JCTC.9.1428,Booth2013}
a novel and promising tool as
demonstrated by the high quality results obtained on Hubbard lattices.  DMET has its
roots in the embedding DMFT framework, where the complexity of the entire
system is reduced by partitioning the problem into a fragment plus an
entangled bath. Together, they constitute an impurity model. As opposed to
DMFT, the DMET impurity model is frequency-independent and therefore
significantly simpler. \cite{PhysRevLett.109.186404}
DMET is designed to reproduce the entanglement of the impurity
rather than its Green's function.
Moreover, the construction of the effective bath is
achieved in an algebraic way. In particular, if the impurity Hamiltonian is
defined in a basis derived from a product state, the complexity of the impurity
basis construction amounts to a rather small matrix
diagonalization. \cite{JPA.39.L85}

Density Matrix Embedding Theory has been benchmarked for model 1D and 2D one-band
Hubbard lattices, \cite{PhysRevLett.109.186404} and in chemical
systems. \cite{JCTC.9.1428}  In the present work, we focus on the former in
order to further investigate the properties of this novel methodology.  In
particular, we analyze the convergence criterion employed in the initial
study. \cite{PhysRevLett.109.186404} We follow the alternative criterion, introduced in Ref.
\onlinecite{JCTC.9.1428} and prove that in periodic systems it allows us to gain
full control over the lattice filling.  Additionally, we investigate the impact
of the effective bath basis on the quality of results. The formalism here presented
allows for spin symmetry breaking in the underlying lattice mean-field
solution that is used to construct the key ingredients of the DMET procedure.
\footnote{Symmetry breaking solutions were presented for the 2D
Hubbard model in Ref. \onlinecite{PhysRevLett.109.186404}, however, a systematic
assessment of spin symmetry breaking was not included}
The only symmetry constraint retained is lattice translational symmetry,
although in the present work we understand it in an extended cell formalism.
This means that the adopted translational unit cell is identical to the DMET
fragment for which the calculations are performed.
We also suggest an approximation that substantially
reduces the number of parameters to be optimized.
In particular, we show that it is sufficient to limit the
density matrix fitting to diagonal elements, i.e., we propose, what we call
\emph{density}, as opposed to \emph{density matrix} embedding. We provide a detailed
numerical analysis of the results obtained with this approach. In particular,
we benchmark the aforementioned approximation against high quality reference
data and exact solutions for energies, two-body correlation functions,
and compressibility.  The latter allows us to access the Mott gap in the
Hubbard model. \cite{PhysRevB.69.195105}
\section{Theory}
The present section is organized as follows. First, for clarity and to make
this work self-contained, we review the basic principles of
DMET. We discuss the properties of the embedding
basis and sketch the algorithm. We then proceed to discuss the
convergence criterion, which is by no means unique. The suggested route
to incorporate broken symmetry embedding is described subsequently.
\subsection{Density Matrix Embedding Theory}

In order to outline the key ideas of DMET, let us assume that we are given the
exact ground state $|\Psi\rangle$ for the system of interest. Then, it is possible to
perform a Schmidt decomposition of this wavefunction according to
\cite{BJP.42.267}
\begin{align}
|\Psi\rangle = \sum_i \lambda_i |\alpha_i\rangle |\beta_i\rangle,
\end{align}
where $|\alpha_i\rangle$ and $|\beta_i\rangle$ can be chosen
such that the former represents a particular set of lattice site states,
which we call the fragment.
The latter must then be the complement that spans
all the sites excluded from $|\alpha_i\rangle$, and we refer to it as
the bath. The summation
in the above equation is limited by the dimension of the smaller
of these two sets. These bases are then used to
project the Hamiltonian into the Schmidt states
of the fragment, $|\alpha_i\rangle$, and bath,
$|\beta_i\rangle$ \cite{PhysRevLett.109.186404}
\begin{align}
\label{Eq:TrueImp}
\hat{H} &\to
        \sum_{ijkl} |\alpha_i\rangle |\beta_j\rangle
                         \langle \alpha_i | \langle \beta_j |
                         \hat{H} | \alpha_{k} \rangle |\beta_l\rangle
                         \langle \alpha_k | \langle \beta_l|  = \hat{H}_{\textrm{imp}}
\end{align}
Details of matrix elements of the projected Hamiltonian,
usually refer to as impurity Hamiltonian, are not relevant here.
For a discussion regarding the impurity Hamiltonian,
the reader is referred to Ref. \onlinecite{PhysRevLett.109.186404}.
The key point to notice is that the size of the new basis (though many-body in principle)
can be chosen much smaller than that of the original problem, if formulated in a
single particle basis.
As discussed in Ref. \onlinecite{PhysRevLett.109.186404}, the original Hamiltonian
$\hat{H}$ and the impurity Hamiltonian $\hat{H}_{\textrm{imp}}$ share the same ground
state $|\Psi\rangle$; information about the expectation values of $\hat{H}$
can be extracted by studying $\hat{H}_{\textrm{imp}}$.

Unfortunately, in practical
applications, we do not have access to the exact solution. The embedding
states have to be approximated. The fundamental simplification of DMET
is to replace the exact solution with a mean-field (here understood as
Hartree-Fock) wavefunction, which is a simple product wavefunction ansatz.
In this case, the Schmidt fragment and bath bases can be represented in terms
of single particle states. This greatly simplifies the computational
treatment of the impurity problem and, as we show in this section, provides an
effective truncation of the dimension of single particle basis for the
problem at hand.

Let us follow the procedure more closely on a specific example,
a Hubbard Hamiltonian. In this case,
\begin{align}
\label{Eq:HubHam}
\hat{H} = \sum_{ij,\sigma} \Big(t_{ij} +v_{ij}\Big)c{}_{i\sigma}^\dagger c_{j\sigma}
+
U \sum_{i} c{}_{i\uparrow}^\dagger c_{i\uparrow} c{}_{i\downarrow}^\dagger c_{i\downarrow}
\end{align}
where $t_{ij}$ connects nearest neighbors, $U$ is the on-site repulsion parameter,
and $v$ is an effective one-body
potential that we wish to introduce; the rationale behind this potential will
become apparent in the following. At this point, let us just notice that $v$ is contained
within the fragment and it is periodically replicated over the lattice.
In the limit of vanishing potential $v$, the Hamiltonian
reduces to the standard Hubbard model.

The first step in DMET is to
solve the above problem at the mean-field level.
This procedure yields a single Slater determinant
$|\Phi\rangle = \Pi_{p}a{}_p^\dagger|0\rangle$,
where $a^\dagger$ creates a hole (occupied) state $|\phi\rangle $
and $|0\rangle$ is a bare vacuum.
The product runs from 1 to the number of electrons.
The hole creation operators are defined by the underlying Hartree-Fock
transformation $\mathbb{D}$ from physical fermions $c^\dagger$,
\cite{ringschuck}
\begin{align}
\label{HartreeFock}
a{}^\dagger_p & = \sum_\mu \mathbb{D}_{\mu p} c{}_\mu^\dagger.
\end{align}
The basis $\mu$ takes into account the position and the spin
degrees of freedom.
In order to construct the fragment and bath states needed to define the
impurity Hamiltonian, we introduce the
operator $\hat{P}_F$ that projects the hole levels onto
the single particle basis contained within the fragment.
Additionally, we have the complement $\hat{P}_B$ such that $\hat{P}_F +
\hat{P}_B = \hat{I}$, where $\hat{I}$ is the identity operator. Following
Ref. \onlinecite{JPA.39.L85}, one can define an overlap matrix
$\mathbb{M}$,
\begin{align}
\label{MMat}
\mathbb{M}_{pq} = \langle \phi_q|\hat{P}_F|\phi_p \rangle,
\end{align}
where the indices $p$ and $q$ run over hole states. This hermitian matrix
can be brought to diagonal form by a unitary transformation $\mathbb{V}$
satisfying $\mathbb{V}^\dagger \mathbb{M} \mathbb{V} = d$, where $d$ contains
at most $\mathrm{min}(n_e,n_f)$ nonzero eigenvalues (here $n_e$ is the number
of electrons in the lattice whereas $n_f$ is the size of the single particle
fragment basis). In the following, we will assume that $n_e \ge n_f$ and that
all $n_f$ eigenvalues are different from 1 and 0 (otherwise, special care has
to be taken while constructing the fragment and bath states).  For each of the nonzero
eigenvalues, one may construct a fragment state
\begin{align}
\label{fState}
|f_i\rangle= \sum_p \frac{\mathbb{V}^*_{pi}}{\sqrt{d_i}} \hat{P}_F |\phi_p\rangle
\end{align}
and a bath state
\begin{align}
\label{bState}
|b_i\rangle = \sum_p \frac{\mathbb{V}^*_{pi}}{\sqrt{1-d_i}} \hat{P}_B |\phi_p\rangle.
\end{align}
The single particle states that correspond to vanishing eigenvalues of
$\mathbb{M}$ are considered as the inert core states $|i\rangle$.  Following
Ref. \onlinecite{BJP.42.267}, the states corresponding to vanishing
eigenvalues of $\mathbb{M}$ denote orbitals with zero probability
of being in the fragment space. Similarly, one can think of
the inert core states as states with vanishing coupling to the
fragment states in the mean-field one-particle density matrix.
The reader is referred to Appendix \ref{Ap:Idem} for more
details.

In DMET, the inert core single particle
states are eliminated from the impurity problem ({\it i.e.} they
are not included while projecting the Hamiltonian
onto the impurity Hamiltonian).
Therefore, DMET retains only a small portion of the
large number of single particle states constituting
the original Hilbert space of the problem.
In fact, the total dimension of the single particle basis
is just twice the dimension of the
single particle basis spanned by the fragment.
This is clear from the particular properties of the
spectrum of $\mathbb{M}$ (Eq. \ref{MMat}) for the mean-field
wavefunction.
In other words, the Hilbert space of the impurity model
is significantly smaller than that of the original problem. It is also
determined by the size of the fragment.

Armed with fragment and bath single particle states, which we shall refer to
as an embedding basis, one can construct an impurity Hamiltonian
(henceforth, for brevity of notation, the
indices in the impurity Hamiltonian denote spin and space coordinates),
\begin{align}
\label{ImpHam}
\hat{H}_{\textrm{imp}} = \sum_{ij} \tilde{t}_{ij} d{}^\dagger_i d_j +
\frac{1}{4} \sum_{ijkl} \tilde{U}_{ijkl} d{}^\dagger_i d{}^\dagger_j d_l d_k
+ \sum_{ij} \tilde{v}_{ij} b{}^\dagger_i b_j .
\end{align}
In the above, $b^\dagger$ denotes bath creation operators whereas $d^\dagger$
denotes either fragment or bath ones. The $\tilde{t}$ and $\tilde{U}$ denote the one- and
(antisymmetrized) two-body
terms of the Hubbard Hamiltonian projected onto the embedding basis, respectively.
Similarly, $\tilde{v}$ corresponds to the additional effective potential.
At this point, we would like to point out that this potential, introduced
in Eq. \ref{Eq:HubHam}, does not affect directly the one-body part of the impurity
Hamiltonian in the fragment space. In other words, the fragment part of the
impurity Hamiltonian corresponds to the physical Hamiltonian.
As the reader may notice, the DMET impurity Hamiltonian
derived from the Schmidt decomposition of the mean-field
wavefunction is expressed in terms of single particle states
and corresponding creation and annihilation operators of fragment
and bath states. This is to be compared with
Eq. \ref{Eq:TrueImp}, which is more general and may include
many-body states as a basis.

The impurity Hamiltonian is the central part of the approximation. As the
dimension of the embedding basis is significantly reduced (by means of
eliminating the inert core $|i\rangle$ states), one is now in a position to
employ powerful ground state computational schemes for solving it.  In
particular, exact diagonalization becomes computationally feasible for modest
fragment sizes.

The goal of DMET is to find an optimal effective
one-body potential $v$ by means of minimizing
\begin{align}
\label{DiffGam}
|\gamma - \gamma_0| =
| \langle \Psi_{\textrm{imp}}|d^\dagger d|\Psi_{\textrm{imp}} \rangle
-
 \langle \Phi|d^\dagger d|\Phi \rangle  | .
\end{align}
In the above, $\gamma$ and $\gamma_0$ are the one-particle density matrix
evaluated with the impurity and the mean-field wavefunction,
respectively. The precise meaning of the convergence criterion is discussed in the
following section.

Once converged, the energy density $e$ for the fragment can be evaluated as
\begin{align}
\label{Eq:EnergyExpr}
e = \sum_{fj} \tilde{t}_{fj} \gamma_{jf}
+
\frac{1}{4} \sum_{fjkl} \tilde{U}_{fjkl} \Gamma_{klfj}
\end{align}
where $\Gamma_{klfj} = \langle d{}^\dagger_f d{}^\dagger_j d_l d_k \rangle$
is the two particle density matrix. The index $f$ in the above summations implies
that at least one of the basis functions has to belong to the fragment space.
Clearly, the energy expression does not correspond to the expectation value taken with
respect to the full lattice Hamiltonian. Because the above expression is not
a true expectation value taken with respect to the full Hamiltonian of the system,
the DMET energy need not be an upper bound to the exact value. In practice,
we have observed a tendency of the procedure to deliver ground state energies
that are below the exact ones.

The self consistency loop in DMET takes the following form
\begin{enumerate}
\item Obtain an initial guess for $v$.
\item Find $\mathbb{D}$ of Eq. \ref{HartreeFock} and
      construct fragment and bath states according to Eqs. \ref{MMat}-\ref{bState}.
\item Construct the impurity Hamiltonian (Eq. \ref{ImpHam}) and solve it.
\item Update $v$ by means of Eq. \ref{DiffGam}.
\item If the update is not negligible, replicate the fragment potential $v$
      over the entire lattice  and go to step 2.
\end{enumerate}

\subsection{Convergence criterion}

DMET provides a very good compromise between accuracy and computational cost
for the Hubbard Hamiltonian. \cite{PhysRevLett.109.186404} However, in the
original formulation, the authors chose to define a convergence criterion based
on an effective one-body potential that minimizes the difference between the
correlated and mean-field one-particle density matrices over the \emph{full} impurity
space, {\it i.e.}, fragment and bath basis. In other words, the effective
potential sought satisfies
\begin{align}
\label{Eq:DMETConO}
\textrm{min}_v \sum_{ij} |\gamma-\gamma_0|_{ij},
\end{align}
where the indices $i$ and $j$ run over the fragment and the  bath
states.  This approach, despite its merits, introduces a certain limitation to
the model. Primarily, as shown in Appendix \ref{Ap:Idem}, because the
mean-field density matrix in the embedding basis must be idempotent, it is not
possible to find $v$  such that the fit between the mean-field ($\gamma_0$) and
correlated one-particle density matrix ($\gamma$) is exact.  An accurate
impurity solver yields $\gamma$ with eigenvalues different from 1 and 0, except
for special cases, like for example, a trivial system of noninteracting
particles. One therefore concludes that in general  $\sum_{ij}
|\gamma-\gamma_0|_{ij} \neq 0$.

This deficiency is not just a formal issue. In fact, since the match
between the density matrices cannot be perfect, the
average number of particles that the impurity Hamiltonian treats
cannot be controlled and can deviate from the desired value.
As clear from Eq. \ref{Eq:EnergyExpr},
the total energies for the physical system computed with DMET
are determined by the fragment energy. It follows then, that any error
in the average number of particles in the fragment affects the predictions
for the entire system. This problem can be ameliorated by changing the
definition of the converged effective one-body potential. An alternative
convergence criterion can be therefore formulated,\cite{JCTC.9.1428}
\begin{align}
\label{Eq:DMETCon}
\textrm{min}_v \sum_{ij\in f} |\gamma-\gamma_0|_{ij}.
\end{align}
Henceforth, the block of the density matrix in the embedding basis
with two indices located on the fragment is referred to as fragment
one-particle density matrix.
With the above definition of the effective potential, provided that the minimum
corresponds to a perfect match, the average number of electrons per
fragment is correct. This statement is particularly
important from the viewpoint of periodic systems, where the average
particle density per fragment is known. The proof of this statement is
presented in Appendix \ref{Ap:Filling}.

The above considerations narrow the choices for the convergence criterion
due to the constraint on the number of particles per fragment. Nonetheless,
fitting only the fragment one-particle density matrix is not a unique choice.
For this reason, our density embedding proposal further
simplifies the numerical procedure while imposing control over the lattice
filling.  As shown in Appendix \ref{Ap:Filling}, an exact match between density
matrices  guarantees that the trace of fragment $\gamma$ will have the desired
number of particles.  Since the trace is obviously determined by the diagonal
elements of the one-body density matrix, we formulate the convergence criterion
as
\begin{align}
\label{Eq:DMETConDiag}
\textrm{min}_{v} \sum_{i \in f} |\gamma-\gamma_0|_{ii}.
\end{align}
The decrease in the number of parameters that needs to be optimized
is accompanied by simplifications of the effective
lattice Hamiltonian that one should solve,
\begin{align}
\hat{H} = \sum_{ij,\sigma} \Big(t_{ij} +v_{ij}\delta_{ij}\Big)c{}_{i\sigma}^\dagger c_{j\sigma}
+
U \sum_{i} c{}_{i\uparrow}^\dagger c_{i\uparrow} c{}_{i\downarrow}^\dagger c_{i\downarrow} .
\end{align}
We find that Eq. \ref{Eq:DMETConDiag} greatly facilitates convergence
while delivering results quantitatively similar to those obtained
with the full method. For the purpose of the present work, we shall denote the results
obtained with the convergence criterion defined by Eq. \ref{Eq:DMETConDiag} as
Density Embedding Theory (DET). In this approach, the effective potential
has a clear physical meaning as an effective (site dependent)
chemical potential.
\subsection{Spin symmetry broken formalism}

Having discussed the convergence criterion in DMET, let us turn our
attention to possibilities for optimizing the embedding basis. The initial DMET
calculations neglected the effect of two-body interactions in the lattice.
\cite{PhysRevLett.109.186404} In other words, the two-body interactions were
suppressed both in the lattice Hamiltonian and bath portions of the impurity Hamiltonian. Only
particles in the fragment states were subjected to the on-site repulsion.  Since in
mean-field the entire system is approximated by a set of non-interacting
particles, the Hartree-Fock transformation constitutes the exact solution; no
symmetry breaking can occur, with some exceptions
(here we do not consider explicit symmetry breaking {\it via} an effective
potential
\footnote{We do not discuss the possibility
of breaking the Hamiltonian symmetry, particularly
the $\hat{S}^2$ symmetry by a spin dependent effective
potential introduced in Eq. \ref{Eq:HubHam}. This is
a possible route for further developments of the model.}).
For example, due to lattice discretization into a set of fragments,
translational symmetry might be violated; this, however, would again correspond
to a situation where one works in an extended unit cell framework.
Another possibility is degeneracy of the solution where the wavefunction
may be chosen to violate certain symmetries of the Hamiltonian.

In the present work, we adopt a different procedure. The full lattice
Hamiltonian corresponds exactly, apart from the effective one-body potential
whose role has been already discussed, to the Hubbard Hamiltonian. The system
is then treated with a spin-unrestricted formalism, where the
Hartree-Fock wavefunction need not be an eigenfunction of the $\hat{S}^2$ operator.
This procedure leads to a spin-dependent embedding basis and an
impurity Hamiltonian that does not need to commute with the $\hat{S}^2$ operator.
In order to retain the simplicity of a spin-restricted formulation,
we do not choose the effective one-body potential to be spin-dependent.
As a direct consequence, we define the effective one-body potential by
\begin{align}
\textrm{min}_v |\gamma^c - \gamma{}_0^c|.
\end{align}
Here, $\gamma_c = 1/2 (\gamma_{\uparrow\uparrow}
+ \gamma_{\downarrow\downarrow})$
is the charge density. The fitted one-particle density matrix may be chosen
either as DMET or DET type.  On the other hand, we note that the broken spin
symmetry formalism does not easily support fitting of the entire one-particle density
matrix in the embedding basis as was done in Ref.
\onlinecite{PhysRevLett.109.186404}.  Let us first stress that the
transformation of the bare fermion basis to the embedding basis is a projection
and hence cannot be inverted (unless, of course, one chooses to divide the
whole system into two equal fragments). Only the transformation of the fragment
states is unitary (and hence invertible). Therefore, solely the $\uparrow$ and
$\downarrow$ fragment density matrices are expressible in the common basis that
defines a charge density matrix. For this reason, fitting of the entire one-particle
density matrix has no clear physical meaning, though it may be numerically performed.
We stress that the arguments outlined above regarding correct filling
are directly applicable to the charge density matrix.  Indeed, in a
spin-restricted formalism, the charge density matrix and the one-particle
density matrix are equivalent.

Finally, let us note that away from half-filling, the spin unrestricted
mean-field solution can admit charge fluctuations that are beyond the size of
the fragment. In order to maintain the mean-field calculations commensurate
with the chosen fragment size, we solve the Hartree-Fock equations in momentum space
assuming homogeneity of the fragment superlattice.
\section{Results and discussion}

In the following sections, we benchmark all the embedding schemes discussed
above for the 1D Hubbard model. For clarity, let us define acronyms that will
be used in the rest of this paper. Results labeled as BA correspond to the
exact Bethe ansatz solution. \cite{PhysRevLett.20.1445} Calculations denoted as
DMET(n) refer to calculations where spin symmetry ($\hat{S}^2$) in the mean-field
solution is allowed to break and fitting of the one-particle density matrix
is performed over the entire fragment chosen to include
n sites.  Similarly, DET(n) denotes
calculations where the fit is enforced only on the diagonal elements of the
fragment one-particle density matrix.  In order to better illustrate the
performance of the broken symmetry approach, we also include data obtained with
embeddings where the two-body interaction is suppressed in the lattice.
To be more precise, in these schemes the Hamiltonian for which the mean-field solution
is obtained  corresponds to Eq. \ref{Eq:HubHam} with $U=0t$, however, the impurity Hamiltonian
does include the on-site interaction but only in the fragment space.
These, as already discussed, correspond to a Non-Interacting case and are denoted as NI
and NI$_F$.  The additional subscript $F$ (for ``Full'' matrix ) implies
fitting of the full impurity one-particle density matrix (this is the method
introduced in Ref. \onlinecite{PhysRevLett.109.186404}).  For convenience, Table
\ref{methods} includes the key qualities of all studied embedding schemes.

\begin{table}[tbp]
\caption{Comparison of the key qualities of all studied embedding schemes.
Eq. \ref{Eq:DMETConO}, Eq. \ref{Eq:DMETCon} and Eq. \ref{Eq:DMETConDiag}
correspond to fitting of full impurity, fragment only, and the diagonal
of the fragment one-particle density matrix, respectively. Two-body
denotes inclusion of these terms in the lattice and the bath
portions of the impurity. Spin Symmetry denotes whether the
mean-field solution is required to be an eigenstate of the $\hat{S}^2$ operator.}
\label{methods}
\begin{tabular}{l c c c c}
\hline\hline
Method   &   Fitting                  & Spin Symm.    & Two-body       & Ref.      \\
DMET     &   Eq. \ref{Eq:DMETCon}     &   NO          &     YES        & This Work \\
DET      &   Eq. \ref{Eq:DMETConDiag} &   NO          &     YES        & This Work \\
NI       &   Eq. \ref{Eq:DMETCon}     &   YES         &     NO         & This Work \\
NI$_F$   &   Eq. \ref{Eq:DMETConO}    &   YES         &     NO         & Ref. \onlinecite{PhysRevLett.109.186404} \\
\hline\hline
\end{tabular}
\end{table}

\subsection{Half-lattice embedding}

\begin{table}[tbp]
\caption{Energy per site (in units of $t$)
for small Hubbard rings at half-filling evaluated with DMET and
DET when the entire lattice is divided into two identical fragments.
Exact values are shown for comparison.}
\label{SmallLattice}
\resizebox{0.95\columnwidth}{!}{
\begin{tabular}{l| c c c c c c}
\hline\hline
 {}& {} & \multicolumn{5}{c}{8 sites} \\
 $U=$      &         &  2        &    4       &    6       &    8       &   10      \\
Exact      &{}       & -0.8210   & -0.5754  & -0.4261  & -0.3333  & -0.2721 \\
DMET(4)    &{}       & -0.8382   & -0.5609  & -0.4118  & -0.3210  & -0.2616 \\
DET(4)     &{} 	     & -0.8210   & -0.5754  & -0.4261  & -0.3333  & -0.2721 \\
\hline\hline
 {} &{}& \multicolumn{5}{c}{4 sites } \\
 $U=$    & {}        &  2        &    4       &    6       &    8       &   10      \\
Exact    & {}        & -0.7071 & -0.5257  & -0.4087  & -0.3301  & -0.2750 \\
DMET(2)  & {}        & -0.7262 & -0.5090  & -0.3860  & -0.3091  & -0.2568 \\
DET(2)   & {}        & -0.7071 & -0.5257  & -0.4087  & -0.3301  & -0.2750 \\
\hline
\hline
\end{tabular}
}
\end{table}

We now focus on a particular benchmark case where the entire lattice
consists of only two fragments. In such a system, at half filling, the Schmidt
decomposition is just a unitary transformation of the bare fermion basis. The
inert core states are absent. Therefore, the complexity of solving the impurity
Hamiltonian is equivalent to solving the original problem.  Results in
Table \ref{SmallLattice} confirms that the DMET embedding scheme
is not exact in this case.
This is because the exact and mean-field one-particle density matrices are not
the same. In particular, they differ in the fragment space.  The DMET equations
are not immediately satisfied and the effective potential has to be optimized.
This leads to a slight but significant deviation of total energies from the
exact ones. The situation is different for the DET scheme. The mean-field
solution charge density does not break translational symmetry and carries
proper filling. This is the case for the exact answer as well. For this reason,
the diagonal elements of both matrices agree. The optimal effective potential
vanishes and the impurity Hamiltonian coincides with the Hubbard model. The DET
embedding scheme converges in one iteration and the computed energy is equal to
the exact one.

\subsection{Hubbard rings at half filling
\label{Sec:HalfFil}}
In order to assess the performance of DMET and DET formalisms, we study the 1D
Hubbard model at half filling. In the present section, the calculations are
performed for a ring of 400 sites with periodic boundary conditions. The exact
solution for half filling was obtained at the thermodynamic limit.
\cite{PhysRevLett.20.1445,
PhysRevA.72.061602,PhysRevA.83.063632,PhysRevLett.90.146402} Errors arising
from finite size effects are negligible for a ring of 400 sites.

\begin{figure}[tbp]
\begin{center}
{\resizebox{\imsize}{!}{\includegraphics{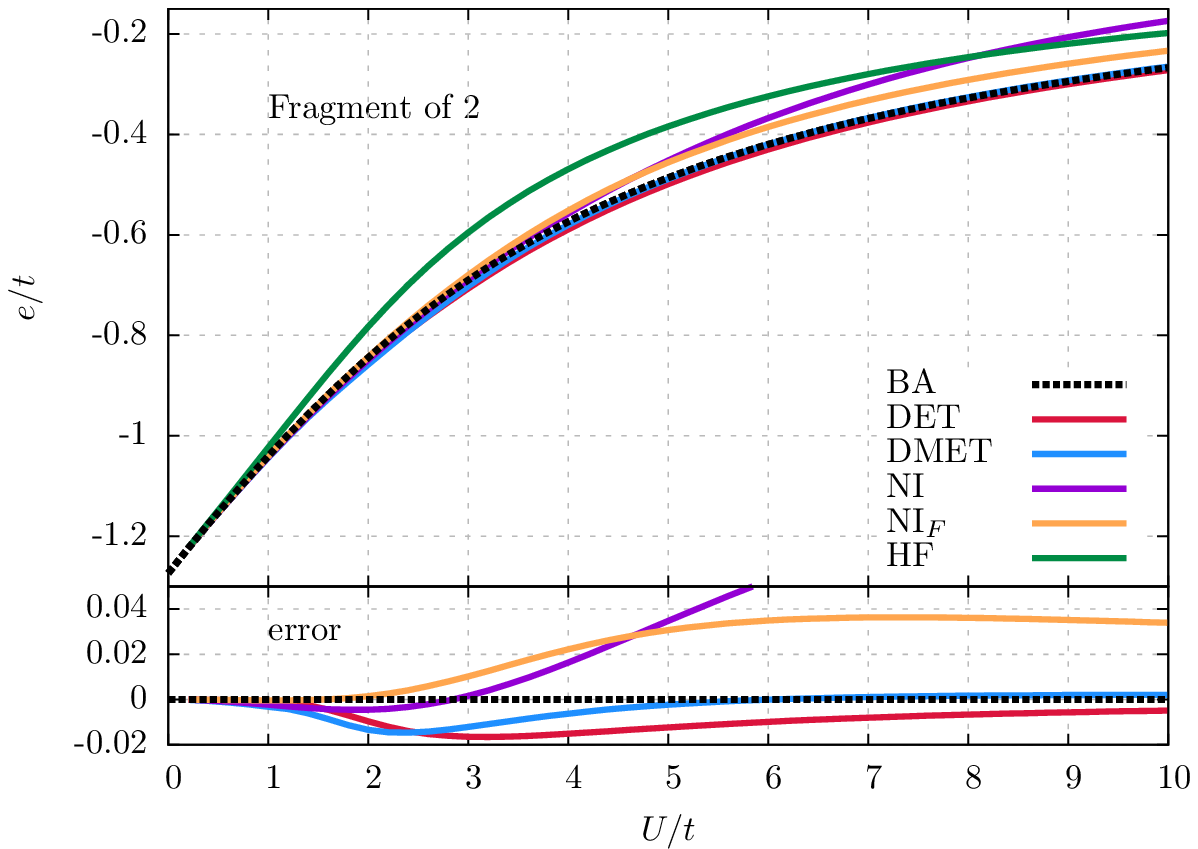}}}
{\resizebox{\imsize}{!}{\includegraphics{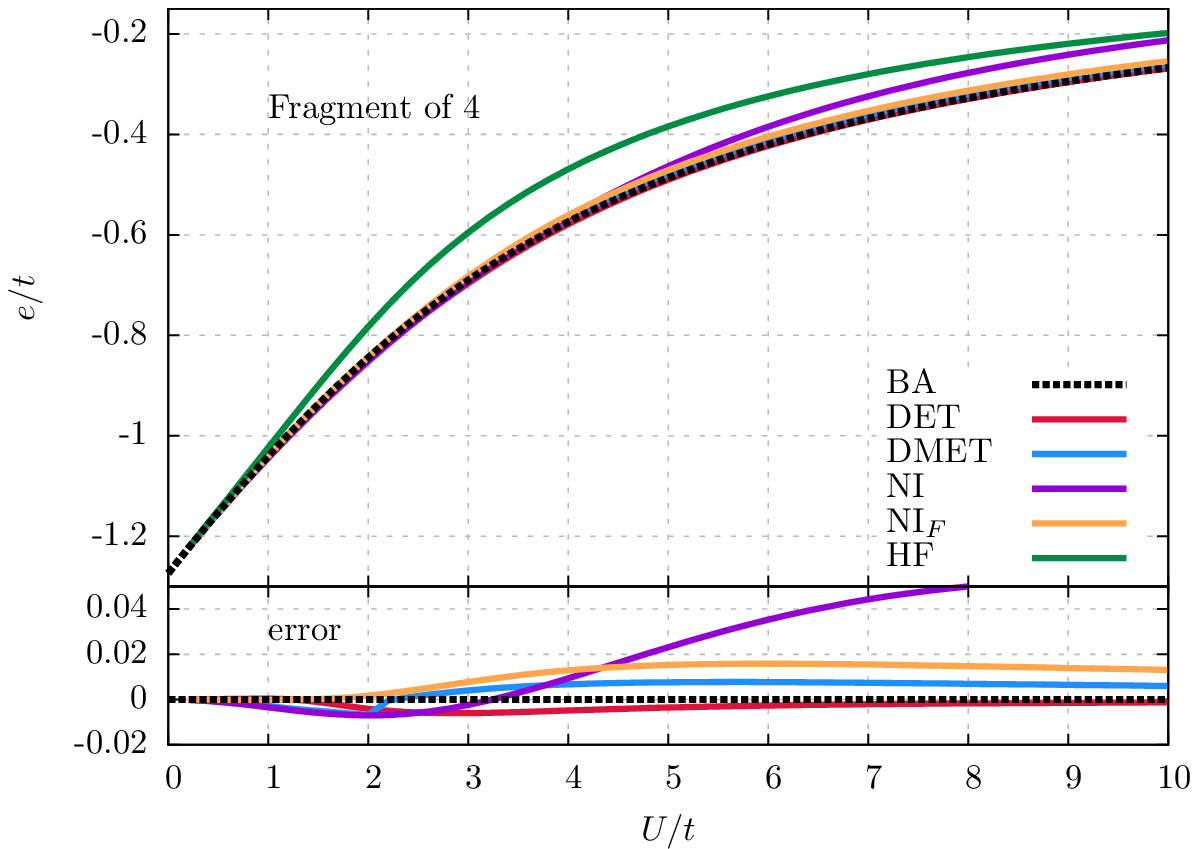}}}
\end{center}
\caption{Energy per site of the 1D Hubbard model at half-filling evaluated
with various embedding approximations with a fragment of 2 (top panel) and
4 sites (bottom panel). Bethe ansatz (BA) and Hartree-Fock (HF) results are added for comparison.
The error with respect to BA is plotted in the bottom panel. The HF wavefunction is not
constrained to preserve $\hat{S}^2$ symmetry.}
\label{HalfEnergy}
\end{figure}

In Fig. \ref{HalfEnergy}, we compare calculations performed with DMET, DET, NI,
and NI$_F$.  The calculations for NI$_F$ were performed using
the program published by Knizia and Chan \cite{PhysRevLett.109.186404} (we have
verified that differences in boundary conditions are negligible by performing
the NI calculations using our current implementation and the one in Ref.
\onlinecite{PhysRevLett.109.186404}).

For a small fragment composed of just two sites, we note that the spin
unrestricted embedding scheme provides already a very good description of the
energy for a broad range of on-site interaction strengths. The difference
between DMET and DET is insignificant within the energy scale of the figure.
This observation supports our choice of determining the optimal fit between the
mean-field and correlated one-particle density matrices.  The performance of
embedding methods based on non-interacting lattice electrons is somewhat worse,
although they still provide an accurate description, especially for smaller $U$
values. The presented data indicates that NI$_F$ is superior to
NI.  Similar results are obtained with a fragment size of 4 sites.  For
this bigger fragment, we observe a systematic improvement of all the embedding
schemes.  In particular, broken spin DET and DMET are virtually
indistinguishable from the exact answer within the scale of the figure. The
NI$_F$ data closely follows the BA curve as well.
Finally, we notice that all the embedding schemes presented
(apart from NI with large value of $U$) constitute a significant
improvement with respect to the Hartree-Fock energy. As shown
in Fig. \ref{HalfEnergy}, Hartree-Fock, even without constraints
to preserve $\hat{S}^2$ symmetry, deviates significantly from
BA, except for $U\to0$ and $U\to\infty$ limits.

In order to gain further insight into the 1D Hubbard half-filled case, we
investigate  the two-body correlation function $\langle n_\uparrow n_\downarrow
\rangle$ (on-site double occupancy). Indeed, a proper description of
energetics combined with accurate double occupancy expectation values implies
that the individual components (one- and two-body contributions) must be
qualitatively correct. The correlation functions for the various embedding
schemes are computed according to
\begin{align}
\langle n_\uparrow n_\downarrow \rangle & = \frac{\langle \hat{U} \rangle}{U n_f},
\end{align}
where $\hat{U}$ is the two-body interaction operator in the fragment
(computed with the impurity wavefunction) and $n_f$ is the fragment size.
For the exact solution, the double occupancy is computed using
the Hellmann-Feynman theorem \cite{PhysRevA.83.063632} as
\begin{align}
\langle n_\uparrow n_\downarrow \rangle & = \frac{\partial e}{\partial U}.
\end{align}

\begin{figure}[tbp]
\begin{center}
{\resizebox{\imsize}{!}{\includegraphics{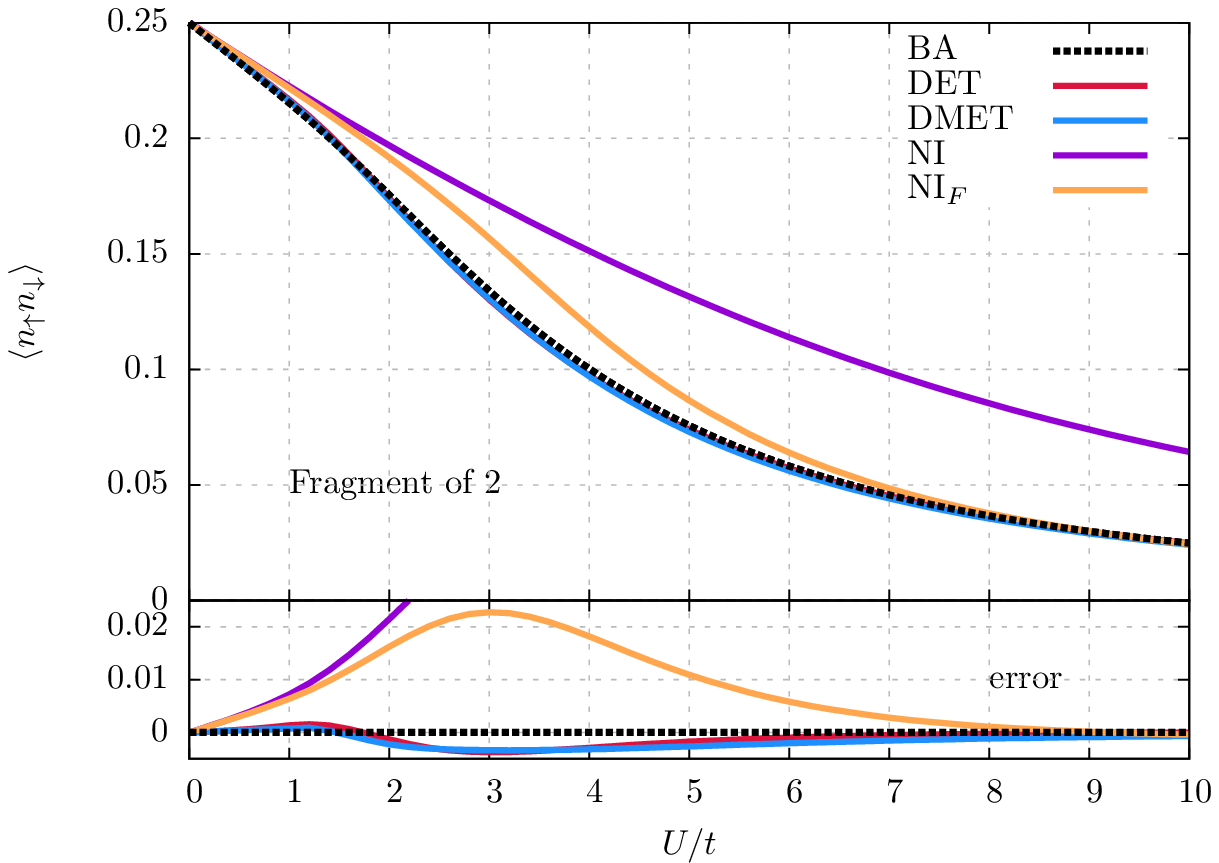}}}
{\resizebox{\imsize}{!}{\includegraphics{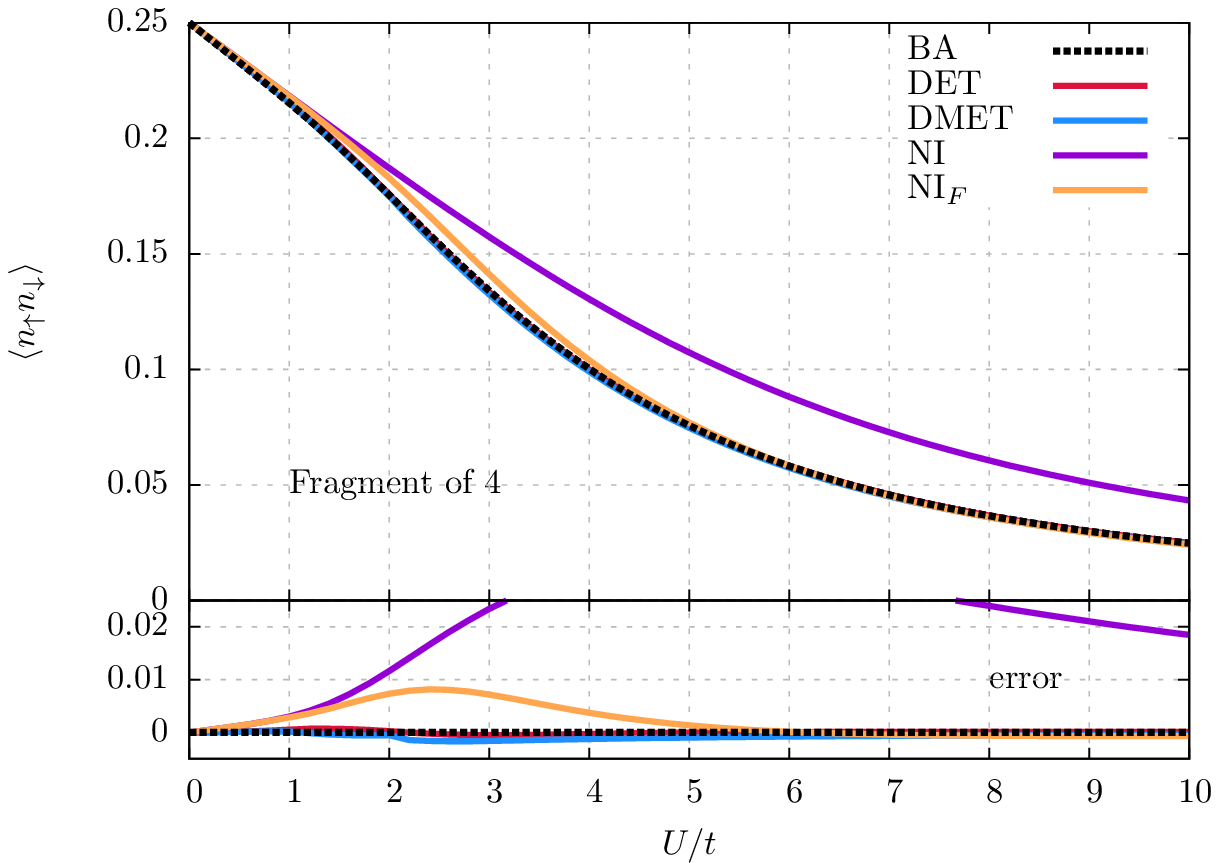}}}
\end{center}
\caption{Comparison of double occupancy for the 1D Hubbard model at
half-filling evaluated with various embedding approximations with a fragment of
2 (top panel) and 4 sites (bottom panel). Bethe ansatz (BA) results are added
for comparison. The error with respect to BA is plotted in the bottom panel.}
\label{DoubleOccupancy}
\end{figure}

Results are presented in Fig. \ref{DoubleOccupancy}.  Once more, the broken
symmetry formalism, even with the smallest fragment, is highly accurate in the
whole range of $U$ values investigated. Just as in the case of the energy,
fitting of the entire fragment density matrix versus its diagonal provides very
similar $\langle n_\uparrow n_\downarrow \rangle$.  This implies that DMET and
DET yield not only similar total energies but also individual components. The
embedding schemes that neglect two-body interactions at the mean-field level
give rise to double occupancy that departs significantly from exact results. In
particular, NI highly overestimates this correlation function.  This in turn
translates into a notably high energy, especially in the strong coupling
regime.  On the other hand, NI$_F$ yields improved behavior of $\langle
n_\uparrow n_\downarrow \rangle$ as a function of $U$ in the strong coupling
regime.  However, the shape of the curve for intermediate couplings reveals
some discrepancies as compared to BA. Increasing the size of the fragment, one
notices an improvement in the trends of double occupancy as a function of $U$.
DMET and DET results are virtually indistinguishable form the exact answer over
the whole studied region. Not much worse are the NI$_F$ results.  Only in the
case of NI, the overestimation of double occupancy in the strong coupling
regime is noticeable. This agrees well with the underestimation of the
correlation energy of the NI approach, which is not ameliorated by increasing
the size of the embedded fragment.

Finally, let us compare the ground state energy density of
DET and DMET with recent variational cluster approach (VCA)
\cite{PhysRevB.77.045133} and cellular dynamical mean-field
theory (CDMFT) \cite{JPCM-21-485602-2009} calculations.
For the on-site interaction $U=4t$ and $U=8t$, the maximum relative
error with respect to Bethe anstatz
for the 2 sites fragment DET or DMET is 2.6\% and 2.0\% respectively,
whereas for 4 sites fragment DET or DMET 1.2\% and 2.1\%.
In comparison, VCA calculations for $U=4t$ become more accurate
than D(M)ET(2) for a cluster size of 6 sites,
and achieve similar accuracy to D(M)ET(4) with a cluster size
of 10. For $U=8t$, D(M)ET is as accurate as the VCA approach
with a cluster of 10 sites. The accuracy of
D(M)ET is also comparable with CDMFT with similar cluster
sizes.

\subsection{Hole-doped Hubbard rings}

\begin{figure}[ht!]
\begin{center}
{\resizebox{\imsize}{!}{\includegraphics{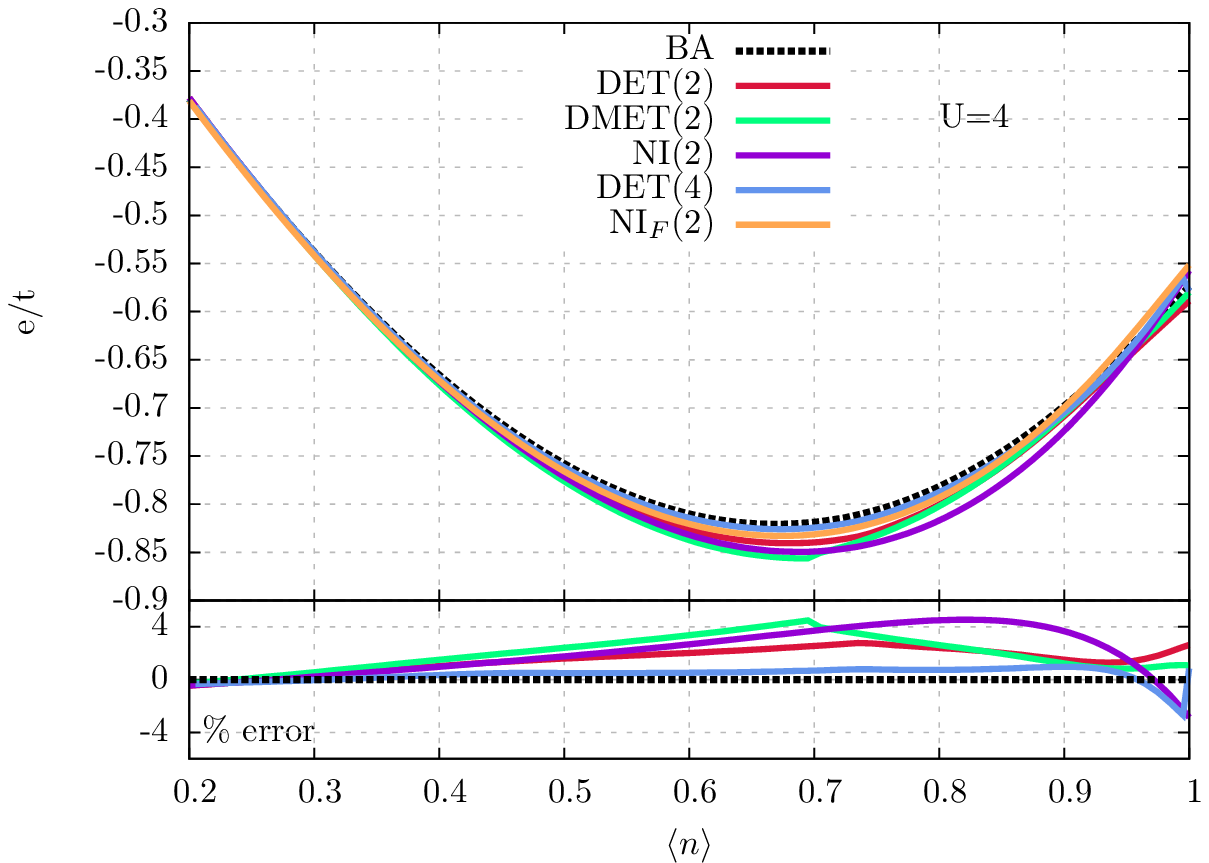}}}
{\resizebox{\imsize}{!}{\includegraphics{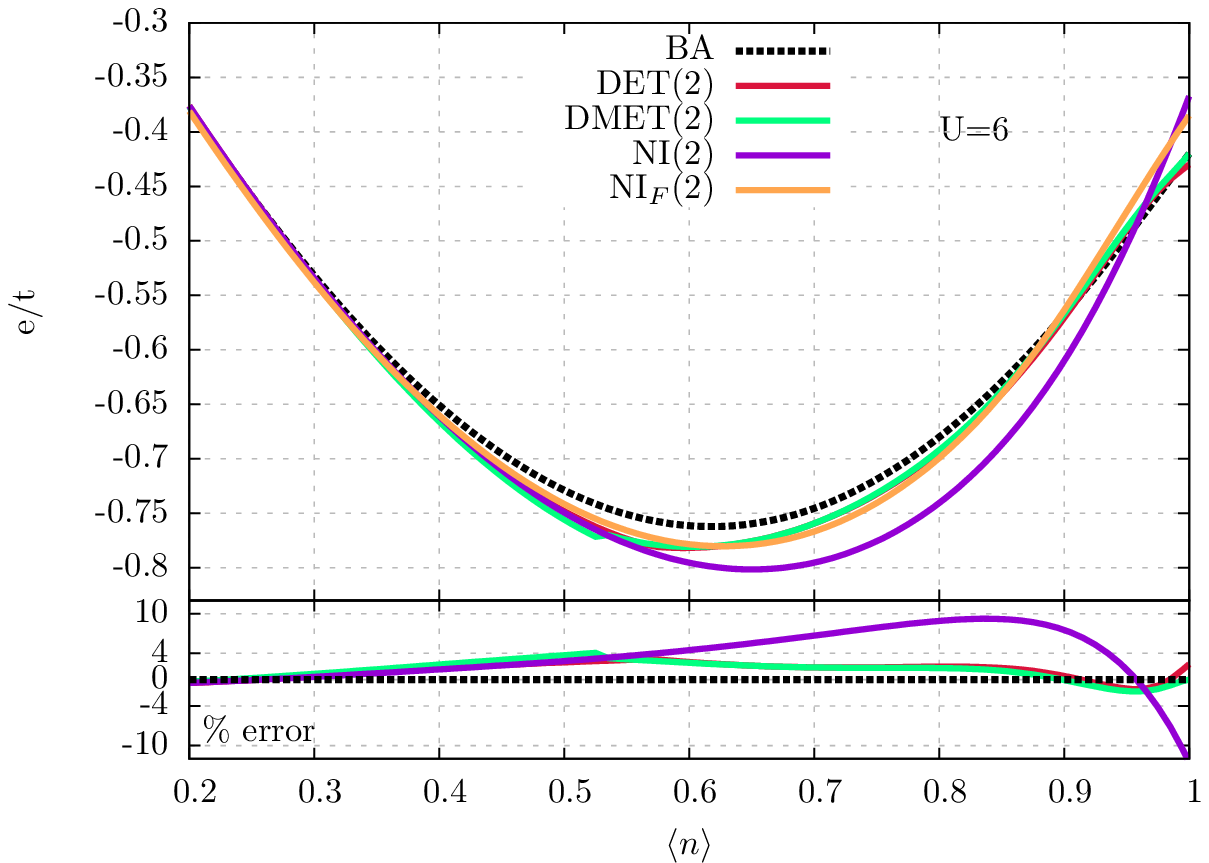}}}
{\resizebox{\imsize}{!}{\includegraphics{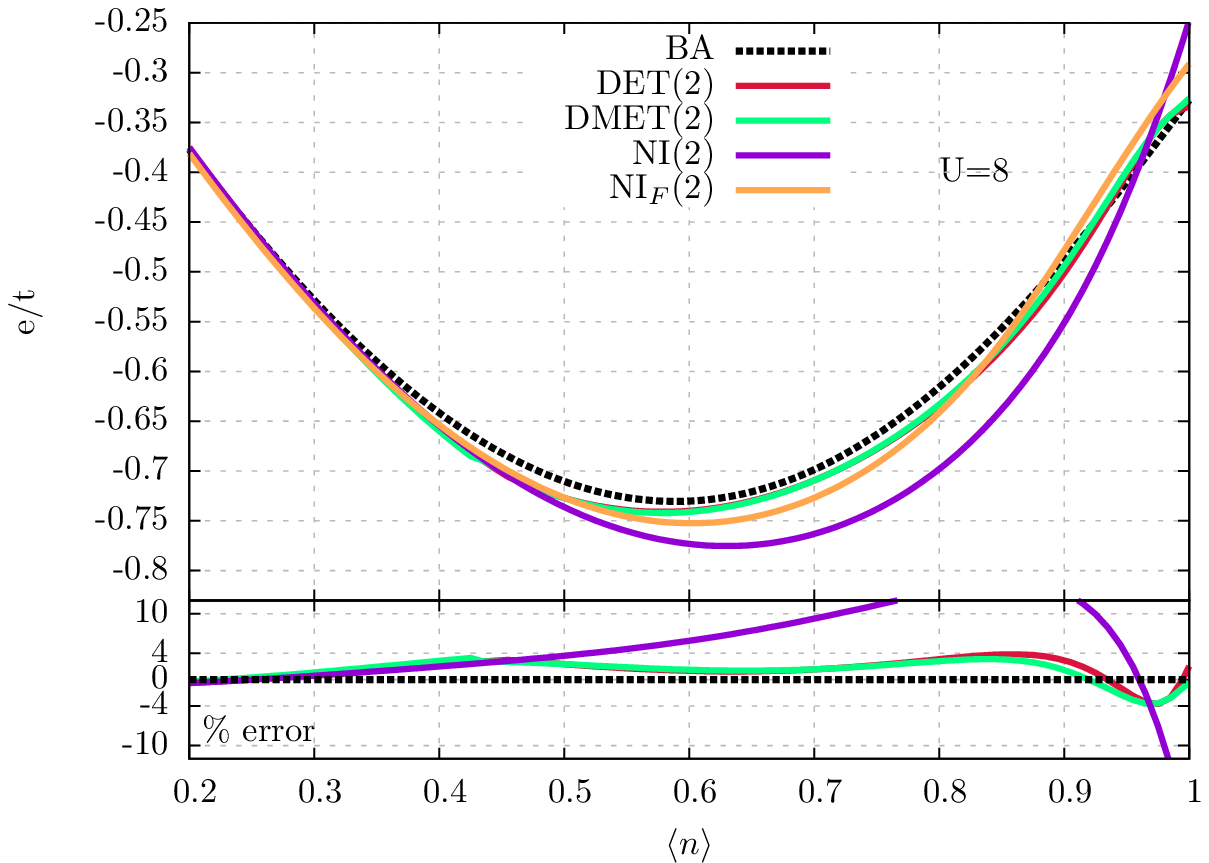}}}
\end{center}
\caption{Energy density of the 1D Hubbard model as a function of hole
doping evaluated with various embedding approximations for $U = 4t$ (top panel),
$U = 6t$ (middle panel), and $U=8t$ (bottom panel). Bethe ansatz
(BA) results are added for comparison. The error with respect to BA is plotted
in the bottom panel.}
\label{EnergyFilling}
\end{figure}

\begin{figure}[ht!]
\begin{center}
{\resizebox{\imsize}{!}{\includegraphics{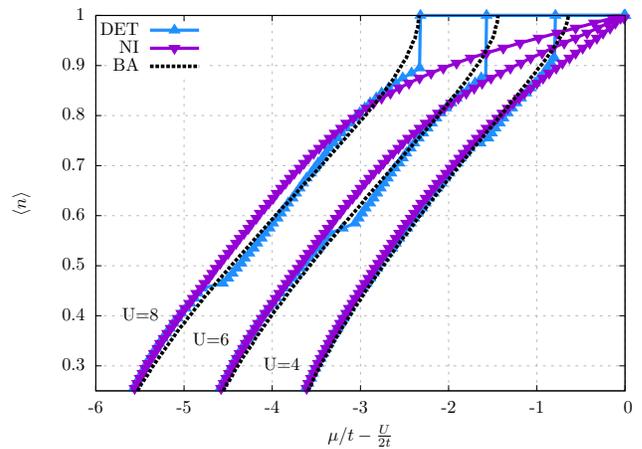}}}
\end{center}
\caption{Lattice filling as a function of chemical potential evaluated
with various embedding approximations. The embedded fragment consists
of 2 sites in each case. Exact Bethe ansatz (BA) values are shown
for comparison.}
\label{Mott-2}
\end{figure}

\begin{figure}[ht!]
\begin{center}
{\resizebox{\imsize}{!}{\includegraphics{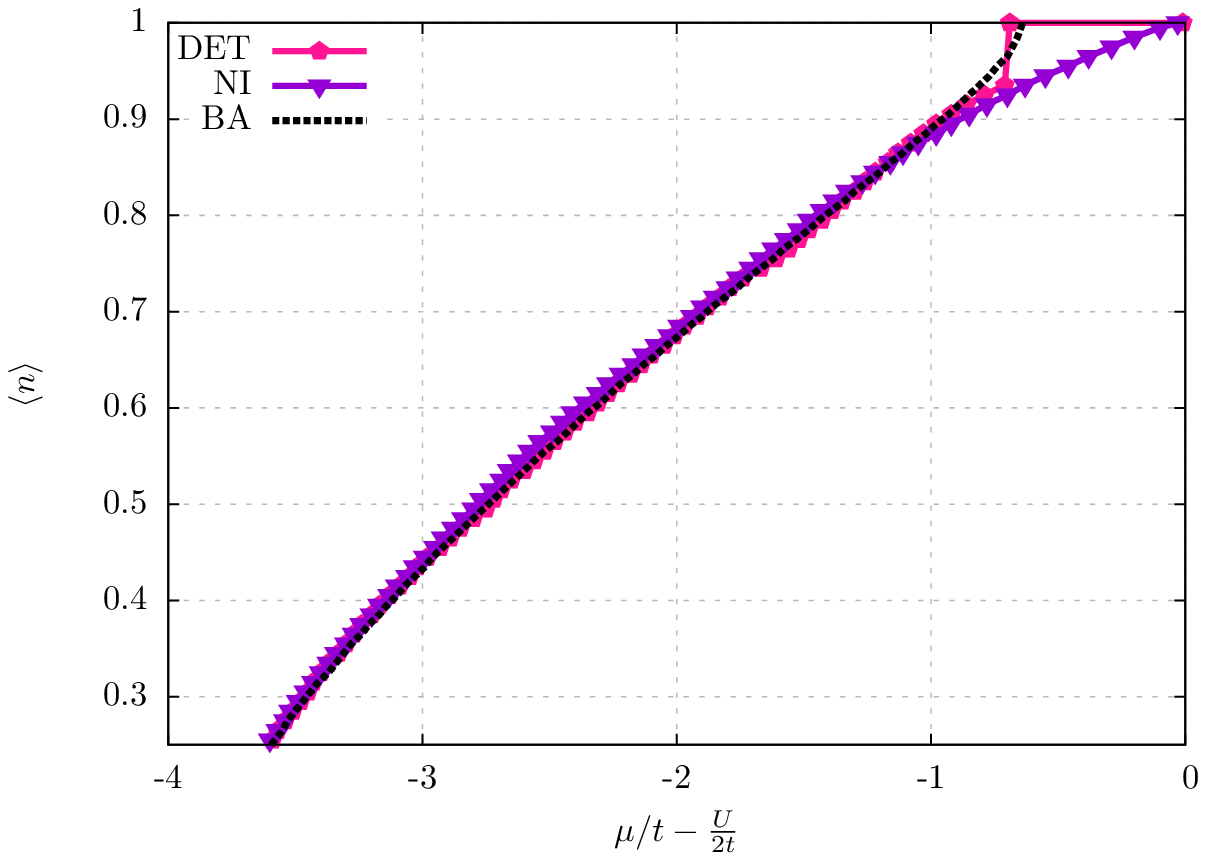}}}
\end{center}
\caption{Comparison of lattice filling as a function of chemical potential
between the DET embedding scheme with a fragment composed of 4 sites
and the exact Bethe ansatz (BA) results. The data corresponds to $U = 4t$.}
\label{Mott-4}
\end{figure}

In this section, we investigate the behavior of the symmetry broken embedding
formalism in the hole-doped Hubbard lattice.  Again, the calculations
correspond to a lattice composed of 400 sites.

In Fig. \ref{EnergyFilling}, we present the dependence of the energy per site
as a function of lattice filling for three values of the on-site interaction, $U
= 4t, 6t$, and $8t$. Additionally, we plot the relative error in the total energy as
compared to exact values. The errors for NI$_F$ are not included because the
data was generated using the program of Ref.
\onlinecite{PhysRevLett.109.186404} where different periodic boundary
conditions were used.  The calculations are performed with a fragment of 2,
except for the $U=4t$ curve evaluated with DET and a fragment of 4. We do not
present a more detailed investigation with larger fragments because of
convergence problems with DMET.  Indeed, we find that inclusion of two-body
interactions in the lattice Hamiltonian introduces significant problems in the
potential optimization; these problems, whatever their origin, do not appear in
the DET density embedding scheme. For instance, even at $U=4t$, we did not succeed
in converging DMET equations for the entire $e(\langle n \rangle)$ curve.

As it is apparent from Fig. \ref{EnergyFilling}, all embedding schemes provide
a fairly accurate description of energetics with the exception of NI. This
embedding scheme quickly breaks down in the strong-coupling limit. Whereas for
$U=4t$ the computed energy does not bear an error of more than around 4\%, the
situation drastically changes for larger on-site interaction strengths. In
particular, the relative error compared to the exact answer easily goes above
10\%. Such a large error is in agreement with the discussion in the previous
section for the half-filed case. The other embedding schemes, NI$_F$, DMET, and DET
significantly ameliorate this deficiency.  For all values of $U$ tested, DET does
not deviate from the BA by more than a few percent. Comparing broken symmetry
methods to NI$_F$, we notice that for $U=4t$, the latter seems to perform
somewhat better, especially for large doping fractions. Nonetheless, results
obtained with the symmetry broken formalism compares favorably with the BA
results.

For larger values of $U$, DMET and DET become more accurate.  Their difference is
mostly pronounced for lattice fillings above 0.9. Indeed, for relatively small
doping fractions and a fragment of 2 sites, the broken symmetry formalism seems
to provide a very accurate description. Finally, let us note that for all
values of $U$ studied, broken symmetry DMET and DET calculations yield very
similar results. Nevertheless, we would like to stress that DET introduces
significant simplifications to the DMET numerical procedure. These
simplifications translate directly into superior convergence for the nonlinear
equations defining the embedding procedure.

Because of the large slope of the $e(\langle n \rangle)$ curve, we would like
to stress that the present DMET and DET schemes have full control over the
lattice filling.  No error does therefore arise from deviations between
electron number per fragment compared to lattice filling.

Increasing the size of the embedded fragment, we observe improvement with
respect to 2 sites for $U=4t$. The correction is most visible for larger doping
fractions where the calculations with smaller fragments yield somewhat
overcorrelated energies. Nonetheless, there is still room for improvement,
especially at small doping fractions. This is the regime where DET(4) differs
from the exact answer by around 3\%. For lattice filling below 0.95, the
relative error drops and the calculations give answers with errors within 1\%.
Finally, let us point out that due to the fact that after a certain doping
fraction, the mean-field no longer breaks spin symmetry, there is a
slight discontinuity in the DMET and DET curves.

Let us now proceed to investigate the lattice density as a function of chemical
potential.  Results are presented in Fig. \ref{Mott-2} and Fig. \ref{Mott-4},
and are obtained by minimizing
\begin{align}
\hat{H} = \hat{H}_0 - \mu \hat{N}
\end{align}
with respect to electron number at a given chemical potential $\mu$.  Here
$\hat{H}_0$ corresponds to the Hubbard Hamiltonian and $\hat{N}$ is the number
operator. In this work, we compare data obtained with the symmetry broken
embedding formalism against the NI one.  Analogous results obtained with NI$_F$
can be found in Ref. \onlinecite{PhysRevLett.109.186404}.
\footnote{To the best of our understanding, the chemical potential for a given
density in Ref. \onlinecite{PhysRevLett.109.186404} was extracted
from the mean-field Fock matrix. In the present work, we do not
use mean-field quantities apart from the embedding basis itself.
In Figs. \ref{Mott-2} and \ref{Mott-4} we do not include results
with NI$_F$ as we do not deem them directly comparable to our results.
We refer the reader to Ref. \onlinecite{PhysRevLett.109.186404} for
a discussion of the NI$_F$ results.}

Whereas the NI$_F$ approach was shown to qualitatively reproduce the Mott gap,
\cite{PhysRevLett.109.186404} we
observe that fitting the fragment only (NI procedure) does not predict a gap.

In the case of DET, the transition is clearly visible. We note however, that
calculations performed with a fragment of 2 sites predict a somewhat abrupt
jump in the $\langle n \rangle (\mu)$ curve around the metal-insulator
transition. While this jump seems like an unfortunate consequence of the
approximations made in the embedding scheme, the actual position of the Mott
transition is rather well reproduced. For $U=4t$ and $U=6t$, the value of the
chemical potential is slightly underestimated compared to the Bethe ansatz
result.  For $U=8t$, DET(2) coincides very well with the exact answer. In the
highly doped regime, the shape of  $\langle n \rangle (\mu)$ is also well
reproduced. The apparent transitions observed at fillings of around 0.75, 0.60,
and 0.45 for $U= 4t$, $6t$, and $8t$, respectively, correspond to filling fractions where
the mean-field calculations no longer break spin symmetry.
This unsatisfactory
behavior does not however fundamentally change the overall good performance of
DET. In Fig. \ref{Mott-4}, we present an analogous $\langle n \rangle
(\mu)$ curve for $U=4t$ obtained with DET and a fragment size of 4. In this case,
one observes improvement of the overall results. The value of the chemical
potential at which the transition occurs, agrees better with the exact results.
Moreover, the description of the highly doped part of the curve is ameliorated.
We note, again, that within the NI scheme, even with 4 sites, the Mott
transition is not observed.

\subsection{Long-range properties}

\begin{figure}[h!]
\begin{center}
{\resizebox{\imsize}{!}{\includegraphics{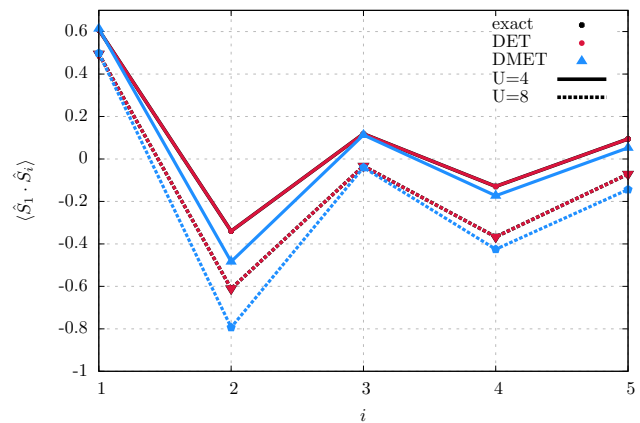}}}
\end{center}
\caption{Spin-spin correlation function evaluated for a Hubbard
ring of 8 sites with DMET and DET embedding. Computations
are performed with a fragment of 4 sites.
DMRG results (deemed exact) are provided for comparison.
For clarity, the results for $U=8t$ are shifted down by 0.2.}
\label{SS-8Sites}
\end{figure}

We now turn our attention to further study two-body correlation functions.
As shown in Section \ref{Sec:HalfFil}, the expectation value $\langle n_\uparrow
n_\downarrow \rangle$ is reproduced very well by DET embedding. No
significant deviations from exact values are observed.  This is not
completely surprising as this two-body correlator is local. DMET is therefore
particularly well suited to compute such local properties.  The question we now wish
to address is whether one could access long-range properties from an impurity
model. To this end, we study the spin-spin correlation function (SSCF),
\begin{align}
{\textrm{SSCF}}(j) = \langle \hat{S}_1 \cdot \hat{S}_j  \rangle
\end{align}
where $\hat{S}_j$ is the spin operator at site $j$.\cite{PhysRevB.87.235129}
We transform this two-body operator into the embedding basis and evaluate it
with the exact solution of the impurity Hamiltonian.  We denote
the first site of the embedded fragment as site 1.  In Fig. \ref{SS-8Sites}, we present
the SSCF evaluated for a lattice of 8 sites and a fragment of 4.  Again, we
notice that in this case the dimension of the impurity problem is equivalent to
the full lattice. The DET scheme is therefore exact.  The DMET optimization
of the effective potential introduces a discrepancy between the predicted and
exact SSCF. Similarly to the energy case, the departure from the exact answer
is small but non-negligible.

\begin{figure}[h!]
\begin{center}
{\resizebox{\imsize}{!}{\includegraphics{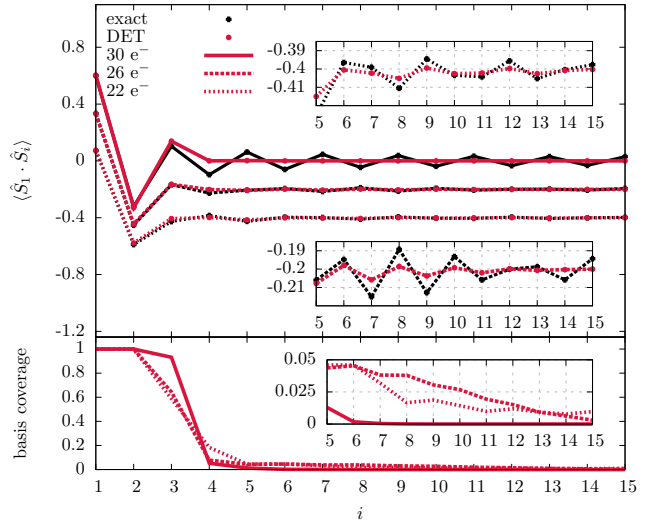}}}
\end{center}
\caption{DET and DMET spin-spin correlation functions (SSF)
evaluated for a ring of 30 sites with 30, 26, and 22 electrons and $U=4t$.
Calculations are performed with a fragment of 2 sites.
DMRG results (deemed exact) are included for reference. For clarity,
data for 26 and 22 electrons is shifted by -0.2 and -0.4, respectively. The
bottom panel presents the basis set coverage (see text for details).}
\label{SS-30Sites}
\end{figure}

Proceeding to systems where the fragment constitutes
only a small fraction of the entire system,
we compute the SSCF for a lattice of 30 sites with 30, 26, and 22 electrons.
Results for the value of on-site interaction $U=4t$ are presented in Fig. \ref{SS-30Sites}.
The agreement between the embedding schemes and
DMRG results is less satisfactory than for other properties.
This discrepancy is expected in embedding schemes where
the impurity Hamiltonian trades the complexity of the
entire lattice by a small fragment connected with bath states.
The bath states, whose number is limited by the size of the fragment
itself, must account for the rest of the lattice that is not included
in the fragment. In order to illustrate this point more clearly, we
plot the partition of the embedding basis single particle states into
distinct sites. For a lattice site $i$, this is
defined as
\begin{align}
\frac{1}{2}\sum_d \langle d | \hat{P}_i | d \rangle
\end{align}
where the summation runs over embedding states. $\hat{P}_i$
is the projector onto site $i$.  The factor of one-half
accounts for the spin degrees of freedom. Clearly, the sites included
in the fragment are completely represented by the impurity Hamiltonian.
The contribution to bath states decays rapidly with distance
from the fragment. This is particularly severe for the half-filled case.
In other words, the embedding basis gets screened and the information
about spatially distant sites is diminished. This results in a very quick
decay of the SSCF. On the other hand, doping the lattice with holes
yields a slower spatial decay. As a consequence,
the SSCF for doped systems has a nontrivial structure. Indeed, we note that
the general trend for this correlator is in agreement with the DMRG reference
values, although the absolute values are underestimated. The problem of
accounting for long range correlations beyond the size of the embedded fragment
is not unique to DMET and also present in DMFT.\cite{PhysRevB.75.045118}

Let us also note that one may consider evaluating long-range properties
by including the inert core states, neglected in the impurity Hamiltonian.
While this may improve the description of long-range order
(especially when the mean-field is qualitatively correct),
we here decided not to include the core states because
they are not an explicit part of the correlated calculations.

\subsection{Spin contamination}

In this section, we investigate the expectation value of $\hat{S}^2$
over the entire lattice.  For the mean-field lattice solution, we evaluate
it explicitly with the mean-field wavefunction.
For the impurity Hamiltonian, we project the lattice
$\hat{S}^2$ operator onto the embedding basis and evaluate it
with the impurity density matrices. The data is shown in Table \ref{SpinCon}.
Here too, we exclude the contribution from the inert core states.

The expectation value of $\hat{S}^2$ is quite similar for DET and DMET.
We also notice a significant decrease
in spin contamination --compared to mean-field values--
when $\hat{S}^2$ is projected onto the embedding
basis and evaluated with the impurity wavefunction.
Finally, let us notice that the lowest filling
beyond which we could no longer obtain a symmetry broken solution
for DMET and DET do not coincide. In particular, for $U=4t$ spin
contamination disappears with DET at a filling of
22 electrons per 30 sites whereas for DMET, we can still obtain a symmetry
broken solution. However, its spin contamination is rather small.

\begin{table}[tbp]
\caption{Expectation values of $\hat{S}^2$ for a ring of 30 sites
evaluated with the mean-field (MF) and the
impurity (imp) wavefunction. The calculations are performed with a fragment
of 2 sites. See text for discussion.}
\label{SpinCon}
\begin{tabular}{l l c c c c c c}
\hline\hline
{}   & {}   &  \multicolumn{6}{c}{NE}                                                           \\
{}   & {}   &  \multicolumn{2}{c}{30}  & \multicolumn{2}{c}{26}   & \multicolumn{2}{c}{22}      \\
{}   &   {} & $S^2_{\textrm{imp}}$ & $S^2_{\textrm{MF}}$ & $S^2_{\textrm{imp}}$ & $S^2_{\textrm{MF}}$ &  $S^2_{\textrm{imp}}$ & $S^2_{\textrm{MF}}$   \\
DMET &($U=4t$) & 0.84        & 7.98       & 0.57        & 4.83       &  0.22        &  1.28        \\
DET  &($U=4t$) & 0.70        & 9.14       & 0.58        & 5.09       &  0.00        &  0.00        \\
DMET &($U=8t$) & 0.98        & 12.3       & 1.06        & 9.65       &  1.09        &  6.80        \\
DET  &($U=8t$) & 0.71        & 13.2       & 1.12        & 10.7       &  1.17        &  7.77        \\
\hline\hline
\end{tabular}
\end{table}

\section{Conclusions}

We have investigated several DMET variants within a broken spin-symmetry formalism
which includes two-body interactions in the lattice mean-field.
We have shown that the numerical procedure can be simplified and its convergence
greatly improved by only fitting the diagonal of the fragment density matrix.
The resulting DET scheme here introduced shows satisfactory accuracy
in Hubbard 1D benchmarks. Additionally, we have demonstrated
that the DET scheme is exact when the half-filled Hubbard lattice is split into
two equal pieces. We deem this property an important guiding principle
for defining a robust embedding approximation.

Our numerical DET data for half-filled lattices is in very good agreement
with exact results obtained with the Bethe ansatz. This
is clear not only for total energies but also for the two-body local correlation
function $\langle n_\uparrow n_\downarrow \rangle$.

For hole doped systems, DET yields good Mott gaps and density profiles.
Moreover, DET retains full control over the average number of particles,
as opposed to the original formulation of Ref. \onlinecite{PhysRevLett.109.186404}.
Both procedures have comparable computational cost except that inclusion of lattice
two-body interactions requires an iterative procedure
for solving the mean-field Hamiltonian.

As currently formulated, DMET cannot accurately describe long-range correlation
functions beyond the fragment size (or, in other words, provide
an accurate description of fluctuations beyond the size of the fragment).
This is an expected shortcoming
inherent to approximations defining an impurity model that is shared
by DMFT. We believe that this deficiency constitutes a major challenge
in the development of DMET.

Further work to improve and address the many shortcomings discussed in this
paper seems warranted. Nonetheless, we believe that DMET is a valuable and
promising quantum embedding tool for studying strongly-correlated systems,
offering high quality results at very low computational cost.

\section{Acknowledgements}
I.W.B would like to acknowledge Dr. Rayner Rodr{\'i}guez-Guzm{\'a}n
for numerous discussions and help with the DMRG calculations.
Dr. Carlos A. Jim{\'e}nez-Hoyos is gratefully acknowledged for
providing the Bethe ansatz data.
This work was supported by the Department of Energy, Office
of Basic Energy Sciences, Grant No. DE-FG02-09ER16053.
G.E.S. is a Welch Foundation Chair (C-0036).
J.D. acknowledges support from the Spanish Ministry of
Economy and Competitiveness under grant FIS2012-34479.

\appendix*
\section{Properties of the Schmidt Basis}
In this Appendix, we prove certain properties of the embedding basis
obtained via the Schmidt decomposition of a
single Slater determinant. Einstein summation convention
is used.

\subsection{Embedding basis}
Let us recall that the embedding basis is obtained from the Schmidt overlap matrix
$\mathbb{M}$
\begin{align}
\mathbb{M}_{pq} = \langle \phi_q|\hat{P}_F|\phi_p \rangle
\end{align}
($p$ and $q$ are the hole states)
which is diagonalized by a unitary matrix $\mathbb{V}$
\begin{align}
\mathbb{V}^\dagger \mathbb{M} \mathbb{V} = d
\end{align}
where $d$ is a diagonal matrix. Let us denote the Hartree-Fock transformation
as $\mathbb{D}$. Then the $i$th fragment and bath states, expressed in
terms of the lattice basis take the form
\begin{subequations}
\label{Ap:Eq:Basis}
\begin{align}
\mathbb{C}{}^F_{\mu i} &=  \frac{\mathbb{V}^*_{pi} \mathbb{D}^F_{\mu p}}{\sqrt{d_i}} \\
\mathbb{C}{}^B_{\mu i} &=  \frac{\mathbb{V}^*_{pi} \mathbb{D}^B_{\mu p}}{\sqrt{1-d_i}},
\end{align}
\end{subequations}
where the superscript $F$ and $B$ denote the fragment and the bath states, respectively.

\subsection{Idempotency of mean-field density matrix in its Schmidt basis}
\label{Ap:Idem}
Let us consider the lattice density matrix $\gamma_0$ that has been obtained
from the mean-field solution. Projected onto the embedding basis, it will take the
following form,
\begin{align}
\label{Ap:Eq:Blocks}
\gamma_0 &=
\begin{pmatrix}
\gamma{}_0^{FF} & \gamma{}_0^{FB} \\
\gamma{}_0^{BF} & \gamma{}_0^{BB} \\
\end{pmatrix}.
\end{align}
Using Eqs. \ref{Ap:Eq:Basis}, for the $FF$ block one obtains,
\begin{align}
(\gamma{}_0^{FF})_{ij} & =
\mathbb{C}{}^{F *}_{\mu i} \mathbb{D}_{\mu r} \mathbb{D}^*_{\nu r} \mathbb{C}{}^{F}_{\nu j} \nonumber \\
& =
\mathbb{V}{}_{pi}
\mathbb{D}{}^{F*}_{\mu p} \mathbb{D}{}_{\mu r} \mathbb{D}{}^*_{\nu r} \mathbb{D}{}^{F}_{\nu q}
\mathbb{V}{}^*_{qj} \frac{1}{\sqrt{d_i d_j}} \nonumber \\
& =
\mathbb{V}{}_{pi} \mathbb{M}{}_{rp} \mathbb{M}{}_{qr} \mathbb{V}{}^*_{qj} \frac{1}{\sqrt{d_i d_j}}
= d_i \delta_{ij}.
\end{align}
In the above, indices $p$, $q$ and $r$ runs over HF hole states while $\mu$ and $\nu$ denote
the on-site lattice spin-orbital.

Analogous straightforward calculations follow for the other blocks. Finally,
\begin{align}
\gamma_0 &=
\begin{pmatrix}
d             & \sqrt{d(1-d)} \\
\sqrt{d(1-d)} & 1 - d         \\
\end{pmatrix}
\end{align}

Similarly, one can easily verify that $\gamma_0$ is idempotent
in the embedding basis and its trace is equal to the dimension
of the fragment single particle basis. On the other
hand, the fragment-fragment block of the density matrix
need not be idempotent. Indeed, keeping in mind that
$ 0 \le d_i \le 1$ (Ref. \onlinecite{JPA.39.L85}) and, as we show
in Appendix \ref{Ap:Filling}, $\sum_i d_i$ is the number
of electrons per fragment, this situation is highly
unlikely. Furthermore, whenever the eigenvalues $d$
are either 0 or 1, one cannot construct an orthonormal
basis according to Eqs. \ref{Ap:Eq:Basis}.

Additionally, let us stress that the inert core states corresponding to
the zero eigenvalues of $\mathbb{M}$ would have vanishing off-diagonal
coupling to the fragment states. This is a consequence of the
orthogonality of the eigenvectors of $\mathbb{M}$.
\subsection{Commutativity of the mean-field density matrix and Fock matrix in the embedding basis}

Let us consider the mean-field Fock matrix in the embedding basis
$f= \mathbb{C}^\dagger \mathbb{F} \mathbb{C}$, where $\mathbb{F}$ is the
lattice Fock matrix. Since $\gamma_0$ and $f$ are Hermitian, they commute if and only if their product,
$t = f \gamma_0$ is Hermitian. Since $t$ and $f$ clearly have the same block structure
as $\gamma_0$ (Eq. \ref{Ap:Eq:Blocks}), one can investigate separately
each block of $t$. And so,
\begin{align}
t{}^{FF}_{ij}
&= f{}^{FF}_{ij} d_j + f{}^{FB}_{ij}\sqrt{d_j(1-d_j)}  \nonumber \\
&= \sqrt{\frac{d_j}{d_i}}
   \Big(\mathbb{V}{}_{pi}
        \mathbb{D}{}^{F*}_{\mu p} \mathbb{F}_{\mu \nu} \mathbb{D}_{\nu q}
        \mathbb{V}{}^*_{qj} \Big)
= \sqrt{d_i d_j} \Big(\mathbb{V}{}^*_{qj} \epsilon_q \mathbb{V}_{qi} \Big)
\end{align}
where $\epsilon$ is the eigenvalue of $\mathbb{F}$.
In the above, one uses the
relation that $\mathbb{D}^F + \mathbb{D}^B = \mathbb{D}$. Similarly,
\begin{align}
t{}^{BB}_{ij} &= \sqrt{(1-d_i)(1-d_j)} \Big(\mathbb{V}{}^*_{qj} \epsilon_q \mathbb{V}_{qi} \Big).
\end{align}
Both of these matrices are manifestly Hermitian. Finally,
\begin{align}
t{}^{FB}_{ij} & =  \sqrt{(1-d_j) d_i} \Big(\mathbb{V}{}^*_{qj} \epsilon_q \mathbb{V}_{qi} \Big)
\end{align}
and
\begin{align}
t{}^{BF}_{ij} & = \sqrt{(1-d_i) d_j} \Big(\mathbb{V}{}^*_{qj} \epsilon_q \mathbb{V}_{qi} \Big)
\end{align}
Clearly $t{}^{* FB}_{ij} = t{}^{BF}_{ji}$, hence $t$ is Hermitian and $\gamma_0$ and $f$ commute.

\subsection{Fragment states can be chosen as bare fermion states}
\label{Ap:Filling}

Let us denote the number of bare fermion single particle states as $M$ and the number
of fragment states as $N$. The fragment basis can be now expressed as
\begin{align}
\mathbb{\tilde{C}}^F &=
\begin{pmatrix}
\mathbb{C}^F \\
0            \\
\end{pmatrix}
\end{align}
which is a $M\times N$ matrix with $\mathbb{C}_F$ being a $N\times N$ matrix.
Since $\mathbb{C}^F$ is a linear transformation
that preserves vectors length, it is unitary,
\begin{align}
\mathbb{C}^{F\dagger} \mathbb{C} = \mathbb{C}\mathbb{C}^{F\dagger} = \mathbb{I}_{N \times N}.
\end{align}
The full embedding basis takes the form
\begin{align}
\mathbb{C}^I =
\begin{pmatrix}
\mathbb{C}^F & 0              \\
 0           & \mathbb{C}^B   \\
\end{pmatrix}
\end{align}
which satisfies $\mathbb{C}^{I \dagger} \mathbb{C}^I = \mathbb{I}_{2N\times 2N}$, but
$\mathbb{C} \mathbb{C}^{I\dagger} \neq \mathbb{I}_{M\times M}$ unless, of course
$N = M/2$. We can now transform the embedding basis with a unitary transformation
\begin{align}
\mathbb{U} &=
\begin{pmatrix}
\mathbb{C}^{F \dagger} & 0          \\
0                      & \mathbb{I} \\
\end{pmatrix},
\end{align}
such that
\begin{align}
\mathbb{C}^{I\prime} = \mathbb{U} \mathbb{C}^I
\end{align}
is expressed in the fragment bare fermion basis. We note
that such unitary transformation does not affect the idempotency
of $\gamma_0$ and its commutativity with $f$.

Finally, let us note that this transformation does not affect the trace of
$\gamma_0$ taken over the $FF$ block. Since the fragment basis is now
equivalent to the bare fermion basis, we see that the trace of
$\gamma{}^F_0$ (hence $\sum_i d_i$) must be equal to the number of electrons
per fragment,
provided that the mean-field
solution does not break translational symmetry (again, assuming that one
works with an extended unit cell chosen as fragment).
%
%
%
%
%
%

\end{document}